\documentclass[
superscriptaddress,aps,prl,reprint,twocolumn,amsmath,amssymb,showpacs]{revtex4-2}

\usepackage{graphicx}  
\usepackage{mathptmx}
\DeclareMathAlphabet{\mathcal}{OMS}{cmsy}{m}{n}
\DeclareSymbolFont{largesymbols}{OMX}{cmex}{m}{n}
\usepackage{epstopdf}
\usepackage{dcolumn}   
\usepackage{bm}        
\usepackage{amssymb}   
\usepackage{amsmath}   
\usepackage{amsthm}
\usepackage{chemarrow}
\usepackage{color}
\usepackage{mathrsfs}
\usepackage{float}
\usepackage{bbold}
\usepackage{bbm}
\usepackage{physics}
\usepackage{comment}

\usepackage[a4paper,colorlinks=true,
linkcolor=blue,citecolor=blue,
pdfauthor={ },
pdftitle={ },
pdfsubject={ },
pdfkeywords={ }]{hyperref}

\theoremstyle{plain}
\newtheorem*{theorem*}{Theorem}

\newcommand{\be}{\begin{equation}}
\newcommand{\ee}{\end{equation}}
\newcommand{\ba}{\begin{eqnarray}}
\newcommand{\ea}{\end{eqnarray}}

\begin{document}


\title{Hunting for Exotic Bosons with Flying Quantum Sensors in Space}

\date{\today}

\author{Xingming Huang}
\email[]{These authors contributed equally to this work}
\affiliation{
CAS Key Laboratory of Microscale Magnetic Resonance and School of Physical Sciences, University of Science and Technology of China, Hefei, Anhui 230026, China}
\affiliation{
CAS Center for Excellence in Quantum Information and Quantum Physics, University of Science and Technology of China, Hefei, Anhui 230026, China}
\affiliation{
\mbox{Hefei National Laboratory, University of Science and Technology of China,
Hefei 230088, China}}

\author{Yuanhong Wang}
\email[]{These authors contributed equally to this work}
\affiliation{
CAS Key Laboratory of Microscale Magnetic Resonance and School of Physical Sciences, University of Science and Technology of China, Hefei, Anhui 230026, China}
\affiliation{
CAS Center for Excellence in Quantum Information and Quantum Physics, University of Science and Technology of China, Hefei, Anhui 230026, China}
\affiliation{
\mbox{Hefei National Laboratory, University of Science and Technology of China,
Hefei 230088, China}}

\author{Xiang Kang}
\affiliation{
CAS Key Laboratory of Microscale Magnetic Resonance and School of Physical Sciences, University of Science and Technology of China, Hefei, Anhui 230026, China}
\affiliation{
CAS Center for Excellence in Quantum Information and Quantum Physics, University of Science and Technology of China, Hefei, Anhui 230026, China}
\affiliation{
\mbox{Hefei National Laboratory, University of Science and Technology of China,
Hefei 230088, China}}

\author{Jiaxi Li}
\affiliation{
CAS Key Laboratory of Microscale Magnetic Resonance and School of Physical Sciences, University of Science and Technology of China, Hefei, Anhui 230026, China}
\affiliation{
CAS Center for Excellence in Quantum Information and Quantum Physics, University of Science and Technology of China, Hefei, Anhui 230026, China}
\affiliation{
\mbox{Hefei National Laboratory, University of Science and Technology of China,
Hefei 230088, China}}

\author{Haowen Su}
\affiliation{
CAS Key Laboratory of Microscale Magnetic Resonance and School of Physical Sciences, University of Science and Technology of China, Hefei, Anhui 230026, China}
\affiliation{
CAS Center for Excellence in Quantum Information and Quantum Physics, University of Science and Technology of China, Hefei, Anhui 230026, China}

\author{Zehao Wang}
\affiliation{
CAS Key Laboratory of Microscale Magnetic Resonance and School of Physical Sciences, University of Science and Technology of China, Hefei, Anhui 230026, China}
\affiliation{
CAS Center for Excellence in Quantum Information and Quantum Physics, University of Science and Technology of China, Hefei, Anhui 230026, China}

\author{Qing Lin}
\affiliation{
State Key Laboratory of Particle Detection and Electronics, University of Science and Technology of China, Hefei 230026, China}
\affiliation{
Deep Space Exploration Laboratory/Department of Modern Physics, University of Science and Technology of China, Hefei 230026, China}

\author{Wenqiang Zheng}
\affiliation{
Zhejiang Provincial Key Laboratory and Collaborative Innovation Center for Quantum Precision Measurement, College of Science, Zhejiang University of Technology, Hangzhou 310023, China}

\author{Yuan Sun}
\affiliation{
Key Laboratory of Quantum Optics, Shanghai Institute of Optics and Fine Mechanics, Chinese Academy of Sciences, Shanghai 201800, China}

\author{Liang Liu}
\affiliation{
Key Laboratory of Quantum Optics, Shanghai Institute of Optics and Fine Mechanics, Chinese Academy of Sciences, Shanghai 201800, China}

\author{Min Jiang}
\email[]{dxjm@ustc.edu.cn}
\affiliation{
CAS Key Laboratory of Microscale Magnetic Resonance and School of Physical Sciences, University of Science and Technology of China, Hefei, Anhui 230026, China}
\affiliation{
CAS Center for Excellence in Quantum Information and Quantum Physics, University of Science and Technology of China, Hefei, Anhui 230026, China}

\author{Xinhua Peng}
\email[]{xhpeng@ustc.edu.cn}
\affiliation{
CAS Key Laboratory of Microscale Magnetic Resonance and School of Physical Sciences, University of Science and Technology of China, Hefei, Anhui 230026, China}
\affiliation{
CAS Center for Excellence in Quantum Information and Quantum Physics, University of Science and Technology of China, Hefei, Anhui 230026, China}
\affiliation{
\mbox{Hefei National Laboratory, University of Science and Technology of China,
Hefei 230088, China}}

\author{Zhengguo Zhao}
\affiliation{
State Key Laboratory of Particle Detection and Electronics, University of Science and Technology of China, Hefei 230026, China}
\affiliation{
Deep Space Exploration Laboratory/Department of Modern Physics, University of Science and Technology of China, Hefei 230026, China}

\author{Jiangfeng Du}
\affiliation{
CAS Key Laboratory of Microscale Magnetic Resonance and School of Physical Sciences, University of Science and Technology of China, Hefei, Anhui 230026, China}
\affiliation{
CAS Center for Excellence in Quantum Information and Quantum Physics, University of Science and Technology of China, Hefei, Anhui 230026, China}
\affiliation{
\mbox{Hefei National Laboratory, University of Science and Technology of China,
Hefei 230088, China}}

\begin{abstract}
Numerous theories beyond the standard model predict the existence of exotic bosons that could serve as candidates for dark matter.
Here present the Space-based Quantum Sensing for Interaction and Exotic Bosons Research Exploration (SQUIRE) scheme, along with a demonstration of a prototype space quantum sensor designed for near-future space experiments.
The core concept involves probing exotic-boson-mediated spin-spin-velocity interactions between the spins within space quantum sensors and the electrons within the Earth.
Unlike terrestrial counterparts, our space-based searches benefit from the significantly increased velocity provided by the orbital motion of the space quantum sensors around the Earth,
which approaches the first cosmic speed.
Additionally, the substantial abundance of polarized electrons within the Earth also enhances the scope of our mission.
We demonstrate that our prototype space quantum sensor can suppress geomagnetic interference in space by 12 orders of magnitude,
{achieving a single-shot sensitivity of 4.3 fT in the sub-mHz regime.}
This illustrates the feasibility of conducting exotic-boson searches in the challenging space environment.
As a result, the search sensitivity for such exotic interactions can be significantly enhanced by up to approximately 7 orders of magnitude compared to both terrestrial experiments and proposals.
This work opens up a novel approach for searching for new physics, including space-based axion-halo searches and CPT violation probes.

\end{abstract}

\maketitle

\textit{Introduction}.--Numerous theories extending beyond the standard model suggest the existence of new bosons,
such as axions and axion-like particles\,\cite{peccei1977cp,wilczek1978problem}, dark photons\,\cite{an2015direct}, paraphotons\,\cite{dobrescu2005massless}, familons\,\cite{ammar2001search}, and majorons\,\cite{dearborn1986astrophysical}.
Many of these bosons are considered potential candidates for dark matter\,\cite{jiangx2021search,duffye2009axions},
which is strongly evidenced to make up most of the mass of the universe.
A key strategy for detecting these bosons involves searching for the exotic interactions they mediate between standard-model fermions\,\cite{moody1984newxet,dobrescu2006spin,fadeev2019revisiting}.
This approach enables the exploration of a wide mass range of exotic bosons without the need for mass scanning in other experiments\,\cite{berger2021prospects,quiskamp2022direct,afach2023can2} and operates independently of the cosmological abundance of dark matter.
Notably,
the exotic interactions include various terms that depend on the relative velocity between interacting fermions, known as spin-spin-velocity interactions (SSVI).
As an example, one of the
SSVI potentials takes the following form:
\begin{equation}
\label{eq:1}
V_{s}=f_{s}\frac{\hbar}{4\pi } \left [ \left ( \hat{\sigma} _{1}\times \hat{\sigma _{2}} \right )\cdot\bm{v}\right]\left ( \frac{1}{r}\right )e^{-r/\lambda }, 
\end{equation}
where $\hat{\sigma } _{i}$ represents the spin
vector of the $i$-th fermion,
$\bm{v}$ is the relative velocity between the particles,
$r$ is the distance between them,
$\lambda =\hbar/m_{\chi}c$ with $m_{\chi}$ being the mass of the bosonic mediator, and $f_{s}$ is the  dimensionless interaction strength of $V_{s}$.
The precision search for such velocity-dependent interactions is fundamentally important.
For example,
by investigating SSVI, 
we can study their mediators, $Z^{'}$ bosons and paraphotons, and explore the parity-time reversal symmetry breaking\,\cite{dobrescu2006spin,fadeev2019revisiting}.

Various terrestrial experiments are conducted in the search for exotic interactions by using a broad range of techniques such as
optical magnetometers\,\cite{kim2019experimental,afach2021search,ji2017searching1,ji2018new},
spin amplifiers\,\cite{wang2023search,wang2022limits,sue2021search,chui1993experimental,arvanitaki2014resonantly,jiang2022floquet,jiang2021floquet,xiong2021searching}, atomic comagnetometers\,\cite{terrano2021comagnetometer,wu2023new,hunter2013using,hunter2014using,clayburn2023}, 
nitrogen-vacancy diamonds\,\cite{chen2021ultrasensitive,liang2023new1,jiao2021experimental},
torsion pendulums\,\cite{ritter1990experimental,terrano2015short,ding2020constraints}, trapped ions\,\cite{wineland1991search,kotler2015constraints}, and other high-sensitivity techniques\,\cite{yan2013new,ficek2017constraints}.
Despite these efforts,
particularly those focusing on studying exotic bosons via SSVI\,\cite{kim2019experimental,jiao2021experimental,ji2018new,sue2021search},
no exotic signals have been detected,
remaining a substantial portion of the parameter space unexplored.
Typically, SSVI experiments involve two polarized objects in relative motion: one serves as a sensitive spin sensor, while the other, acting as the spin source, produces the exotic field to be measured.
To maximize potential exotic signals,
objects with high polarization number are preferred as spin sources,
which are then moved at high velocities relative to the spin sensor.
However, there is a fundamental trade-off between increasing the polarized spin number in the spin source and simultaneously enhancing its velocity,
setting a significant bottleneck in enhancing the interaction amplitudes.
For example, when a spin source contains $10^{25}$ polarized electrons,
its velocity is typically limited to about 20\,m/s\,[\onlinecite{ji2018new}],
severely constraining search sensitivities.
Overcoming this limitation requires developing innovative methods for exotic-boson searches.

In this Letter, 
we introduce a Space-based Quantum Sensing for Interaction and Exotic Bosons Research Exploration (SQUIRE) scheme and develop a prototype space quantum sensor to provide substantial support for this space-based scheme.
The core concept leverages the extremely high velocity of space platforms deployed with spin-based quantum sensors, which can reach up to 7.67 km/s—close to the first cosmic speed and two orders of magnitude higher than previous experimental velocities\,\cite{kim2019experimental,ji2018new,sue2021search}.
Additionally, the Earth naturally generates polarized geoelectrons, yielding about $10^{17}$ times more polarized particles than laboratory spin sources.
By combining these advantages, our scheme enhances the SSVI search sensitivity surpassing terrestrial works by up to 7 orders of magnitude.
To design the search scheme, we numerically simulate the essential characteristics of exotic interactions in the space-based scheme, including their orientations, amplitudes, and frequency spectra.
To validate the feasibility of exotic-boson searches in the challenging space environment,
we experimentally demonstrate a prototype space quantum sensor with the ability to suppress the interference from temporally-varying geomagnetic field in space by 12 orders of magnitude while achieving a high single-measurement sensitivity of 4.3\,fT at frequencies below 1\,mHz.
Using this sensor, our scheme can improve the search sensitivity for all SSVI in the force range
$\lambda>10^{6}\,\mathrm{m}$, achieving at least six-order-of-magnitude improvements for $V_{8}$, $V_{14}$, and $V_{16}$ compared to previous terrestrial works\,[\onlinecite{ji2017searching1},\onlinecite{hunter2014using}].
In addition to SSVI, our scheme also achieves significant improvements for spin-spin interactions and spin-mass-velocity interactions\,\cite{ji2017searching1,hunter2014using,clayburn2023,wu2023new}.
This work would open a door to a broad range of investigations in fundamental physics, such as searches for the axion halo surrounding the Earth\,\cite{banerjee2020searching}.


\textit{Space-based search scheme}.--We show that numerous existing space platforms offer unparalleled experimental conditions for equipping quantum sensors and performing SSVI searches\,\cite{tsai2023direct,liu2018orbit1,bassie2022way,gu2022china,gibney2015dark,abdo2008fermi,bergstrom2013new}, particularly in terms of their high velocities. 
For example, the International Space Station, the China Space Station, medium Earth orbit satellites, and geostationary satellites exhibit velocities of up to 7.67\,km/s, 7.67\,km/s, 4.5\,km/s, and 3.1\,km/s, respectively.
Among these, space stations are particularly advantageous, with velocities that surpass those achieved in terrestrial experiments by at least two orders of magnitude\,[\onlinecite{kim2019experimental},\onlinecite{ji2017searching1},\onlinecite{sue2021search}].
As a representative platform, the China Space Station, which was completed in 2022\,[\onlinecite{gu2022china}], is chosen to discuss our SQUIRE scheme in the following sections.

\begin{figure}[t]
    \centering
    \includegraphics[scale=0.153]{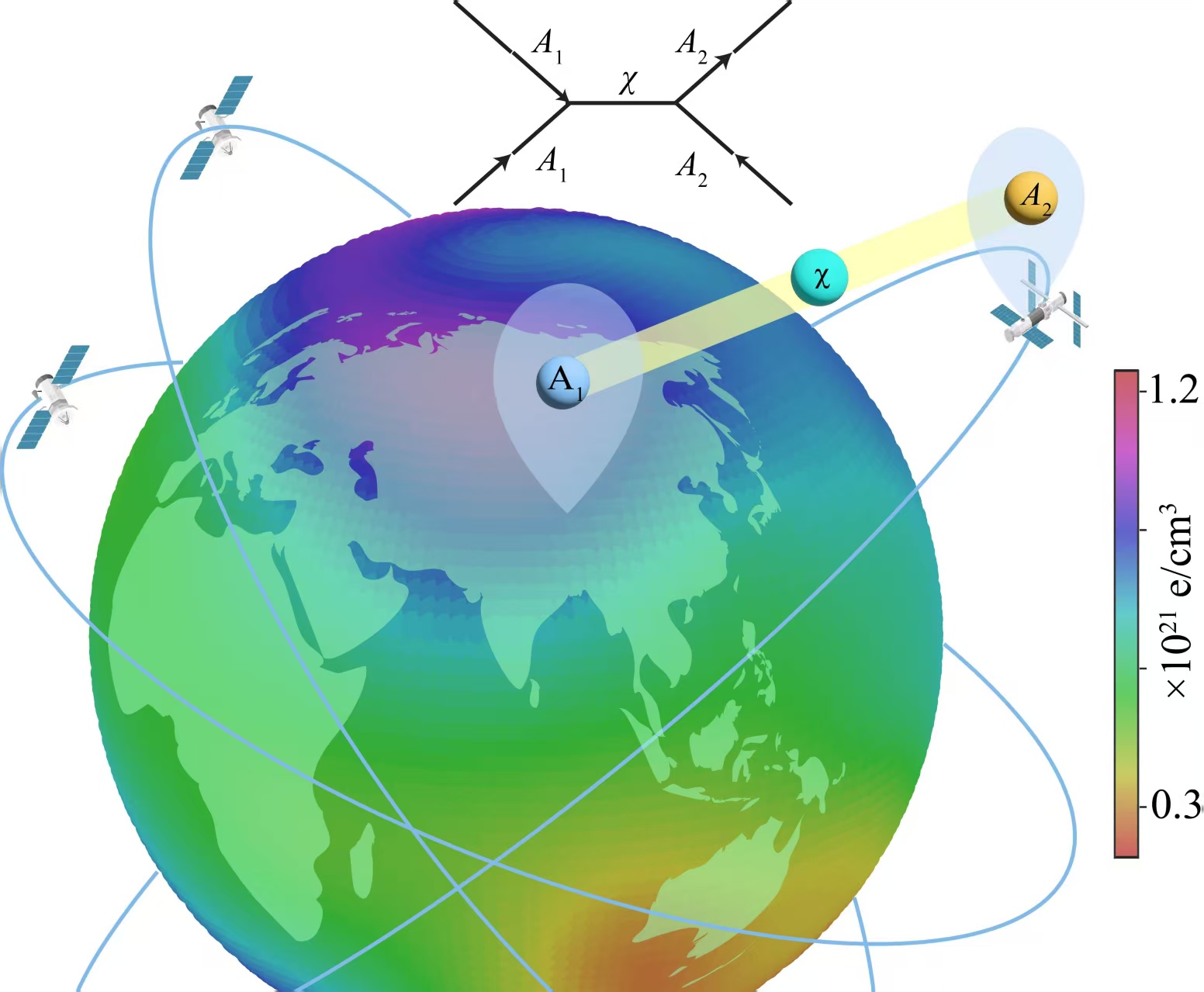}
    \caption{Concept of the space-based searches for exotic bosons.   
   To enhance the exotic interactions, our scheme leverages the extremely high velocity of space platforms deployed with highly sensitive quantum spin sensors.
   In the searches, flying quantum sensors are used to search for the exotic field generated by the abundant polarized geoelectrons within the Earth.
   The distribution of polarized geoelectrons
   is numerically simulated and presented in this figure.
   We illustrate the distribution of polarized geoelectrons on a spherical surface centered at the Earth's core, with a radius of 5000\,km.
   The color shading represents the magnitude of the polarized geoelectron-spin density.   } 
    \label{fig:figure 1}
\end{figure}

We consider the spin-spin-velocity interactions between the spins within a space quantum sensor and the polarized geoelectrons within the Earth (as spin source) (Fig.\,\ref{fig:figure 1}).
Similar to the Zeeman interaction, the effect of an SSVI interaction can be viewed as the polarized electrons within the Earth generating a pseudomagnetic field on the space quantum sensor.
The pseudomagnetic field $\bm{B}_{\mathrm{pseu}}$ can be expressed as
\begin{flalign}
\label{eq:2}
\bm{B}_{\mathrm{pseu}}=-\int_{V}^{}\rho \left ( \bm{r}  \right ) \frac{f_{s}\hbar}{4\pi  \mu _{N} } \left ( \hat{\sigma}_{1}\times \bm{v} \right )\left ( \frac{1}{r}\right )e^{-r/\lambda }d\bm{r},
\end{flalign}
where $\hat{\sigma}_1$ and $\rho \left ( \bm{r}  \right )$ represents the polarization direction and polarization density of geoelectrons at position $\bm{r}$, $V$ represents the integration volume of geoelectrons,
and $ \mu _{N}$ is nuclear magneton, considering the nucleons within the spin senor that participate in the exotic interactions.
In order to numerically simulate the pseudomagnetic fields, the essential parameters, $r\left ( t \right ) $, $\bm{v}\left ( t \right ) $, and $\rho \left ( \bm{r}  \right )$, must be determined.
Among them, $r\left ( t \right )$ is related to the orbital revolution of the space station and can be acquired by using its two-line element (TLE) data, while $\bm{v}\left ( t \right ) $ can be derived from the time derivative of $r\left ( t \right ) $
Here we use the two-line element (TLE) data of the China Space Station from May 20, 2022\,\cite{spacesd2} to illustrate the space-based scheme.
The data includes four primary parameters: the revolution-orbit radius ($R$) of 6700 km, the mean motion ($n$) of 15.61, the orbital inclination ($\theta$) of 41.45 degrees, and the right ascension of the ascending node ($\phi$) of 84.98 degrees.

The distribution of polarized geoelectron-spin density $\rho \left ( \bm{r}  \right )$ is a critical parameter to estimate pseudomagnetic fields. 
Polarized geoelectrons result from the inverse polarization of unpaired geoelectrons within the geomagnetic field\,\cite{hunter2013using},
and their density is expressed as
\begin{flalign}
\label{eq:3.0}
\rho \left ( \bm{r}  \right ) = \rho_{e} \left ( \bm{r}  \right )\frac{2\,\mu _{\mathrm{B}}\,B_{e}(\bm{r}) }{k_{\mathrm{B}}\,T(\bm{r})}, \end{flalign}
where $\mu _{\mathrm{B}}$ is the Bohr magneton, $k _{\mathrm{B}}$ is Boltzmann’s constant, $T\left ( \bm{r}  \right )$ and $\rho _{e} \left ( \bm{r}  \right )$ represent the temperature and the density of “equivalent $\mathrm{Fe}^{2+}$\,$e^{-}$" at position $\bm{r}$, respectively.
The value of $\rho _{e} \left ( \bm{r}  \right )$ can be derived from data in fields such as deep-Earth mineral physics, geochemistry, and seismology\,\cite{hunter2013using}.
The geomagnetic field $B_{e}(\bm{r})$  is derived from a magnetic potential
approximated by a 12th-order associated Legendre expansion in the World Magnetic Model\,\cite{mausi2010us,buffett2000earth}.
Figure\,\ref{fig:figure 1} illustrates the numerical results of the distribution of polarized geoelectron-spin density, depicted through the color overlay on the Earth.
These results clearly reveal a distinct non-uniformity.
The total number of polarized geoelectrons in our simulations exceeds $10^{42}$, highlighting the immense scale of the spin source in the space-based scheme.

\begin{figure}[t]
    \centering
    \includegraphics[scale=0.65]{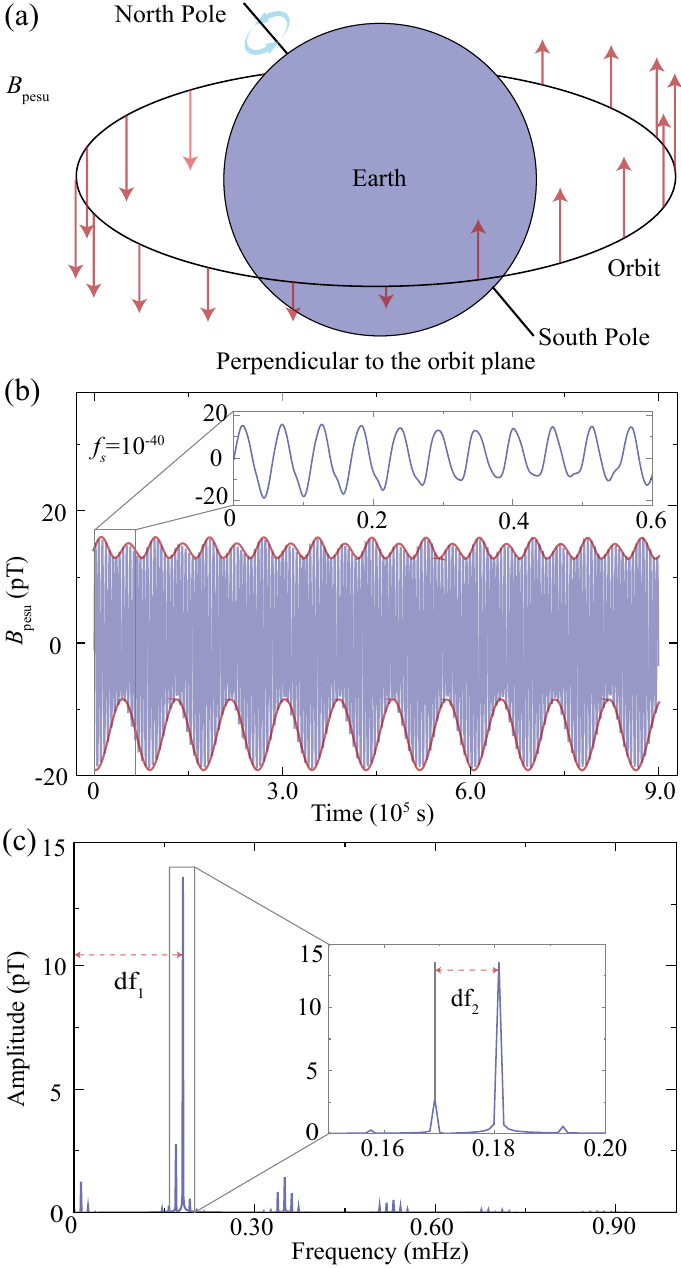}
    \caption{Results of the simulated pseudomagnetic signal on the China Space Station from May 20, 2022, to June 1, 2022.
    (a) Orientation for the oscillating component of $\bm{B}_{\mathrm{pseu}}$ in various locations of the orbit of the China Space Station.
    We find that the optimal detection direction for $\bm{B}_{\mathrm{pseu}}$ is perpendicular to the orbital plane. 
    (b) Temporal variation of the simulated signal of $\bm{B}_{\mathrm{pseu}}$ on the orbit. 
    The signal is the projection of the $\bm{B}_{\mathrm{pseu}}$ in its optimal detection direction. 
    (c) Frequency spectra for ${B}_{\mathrm{pseu}}$.
    Its primary frequency corresponds to the space-station revolution at $\mathrm{df_1} \approx 0.189$\,mHz.
    The split of the main peak with a frequency difference of $\mathrm{df_2}\approx 0.012$\,mHz is generated by the rotation of the Earth. 
    Moreover, the asymmetry of polarized geoelectron-spin distribution gives rise to their weaker high-frequency harmonic peaks.}
    \label{fig:figure 3}
\end{figure}

We conduct an in-depth analysis of the characteristics of pseudomagnetic fields $\bm{B}_{\mathrm{pseu}}$,
which could potentially be detected by a spin sensor aboard the China Space Station.
This analysis includes the examination of their orientations, amplitudes, and frequency spectra, as depicted in Fig.\,\ref{fig:figure 3}.
Our findings indicate that $\bm{B}_{\mathrm{pseu}}$ predominantly aligns perpendicularly to the orbital plane and features an amplitude that oscillates periodically with a period of $\mathcal{T}\approx1.5$\,hours (Fig.\,\ref{fig:figure 3}a).
To optimally search for the SSVI $V_s$,
it is recommended that the sensitive axis of the quantum sensors in orbit be oriented perpendicular to the orbital plane.
Considering the most stringent constraints on $f_{s}$ from terrestrial experiments\,\cite{hunter2014using},
our scheme predicts that the amplitude of $\bm{B}_{\mathrm{pseu}}$ can reach a significant 20\,pT (Fig.\,\ref{fig:figure 3}b), which is well within the detection capabilities of existing spin-based quantum sensors\,\cite{jiangx2021search,kominis2003subfemtotesla,walker2016spin}.
In comparison, the amplitude of $\bm{B}_{\mathrm{pseu}}$ in terrestrial experiments is only approximately 0.015\,pT\,\cite{hunter2013using,hunter2014using}, highlighting the advantage of the space-based approach.
Furthermore, the motion of the space station modulates the exotic signals—typically appearing as DC signals in terrestrial experiments\,\cite{hunter2014using}-into periodic signals, with the primary frequency corresponding to the station's orbital revolution at $\mathrm{df_1=1/\mathcal{T}}  \approx 0.189$\,mHz (Fig.\,\ref{fig:figure 3}c).
The periodic nature of the signals facilitates their extraction from background noise, significantly enhancing the accuracy and sensitivity of the experiment.
Despite the ultra-low frequency of these exotic signals, we demonstrate the feasibility of detecting them in subsequent discussions.



\textit{{Prototype} space quantum sensor}.--Despite the significant enhancement of pseudomagnetic signals achieved by the space-based scheme, the complex space environment introduces substantial noise not present in terrestrial laboratory experiments.
{To implement the space scheme, the space quantum sensor must suppress the space environmental noise while maintaining high sensitivity, long-term stability, and the finite volume required by space station payload limitations\,\cite{spacesd3}, which poses significant challenges for its prototype design as discussed later.}
Here we first outline the primary challenges:

(i) {Interference from the geomagnetic field:}
As the space station orbits, the flying quantum sensor experiences a temporally varying signal $B_{\mathrm{gmf}}$ induced by the geomagnetic field, with an amplitude of approximately 20\,$\mu$T in our simulations. 
This signal shares nearly identical frequency characteristics with the exotic signals, severely interfering with the space-based searches.
To meet the sensor's requirements for a 100-day mission, $B_{\mathrm{gmf}}$ must be suppressed by 12 orders of magnitude, from 20\,$\mu$T to 0.02\,fT, to ensure it falls below the search sensitivity threshold.
While various types of magnetic shields are developed, as analyzed in \cite{thomas1968magnetic}, common solutions achieve suppression of only 5 to 6 orders of magnitude \cite{kelha1982design},
highlighting the need for innovative approaches to bridge this gap.

(ii) {Noise from space station vibrations:}
The vibration of space station induces a rotation in the coordinate system of the space quantum sensor,
denoted as $\Omega_{\rm{rot}}$,
leading to an effective shift in the atomic energy levels. 
The magnitude of $\Omega_{\rm{rot}}$ is approximately 0.005$\,^{\circ} /\mathrm{s}$\,\cite{spacesd3}. 
In our sensor, this corresponds to an equivalent magnetic noise of 1.9\,pT, which can severely reduce the space-based search sensitivity.

(iii) {Disruptions caused by cosmic radiation:}
Cosmic radiation can induce data errors in the registers of control circuits, causing disruptions in the operation of spin sensors\,\cite{scheick2019guideline,sosmicradiations1}.
For commercial sensors without radiation resistance, cosmic radiation in the space environment could cause an average of 30 disruptions per day\,\cite{scheick2019guideline} 
This could lead to approximately two disruptions within the 1.5-hour exotic signal period, which is unacceptable for the searches.

To address the primary challenge—the interference noise $B_{\mathrm{gmf}}$—the shielding capability of the magnetic shield must be greatly improved.
To achieve this, we experimentally demonstrate a magnetic shield as a five-layer cylindrical structure, with the diameters of the layers increasing geometrically by a common ratio of about 2 from the inner to the outer layer.
In addition to its optimized structure, the magnetic shield utilizes $\mu$-metal materials with high magnetic permeability. 
As a result, even in a compact form with the dimensions of
$24\times 40\times 24$\,$\mathrm{cm^{3}}$, the magnetic shield can suppress $B_{\mathrm{gmf}}$ by 8 orders of magnitude at frequencies below 1\,Hz in experimental tests, as detailed in the Supplemental Material\,\cite{sup1}.

To achieve the additional four orders of magnitude suppression necessary to complete the total required 12 orders of magnitude reduction of $B_{\mathrm{gmf}}$,
we experimentally demonstrate a dual noble-gas spin sensor\,\cite{gao2024stability}.
By using the opposite signs of gyromagnetic ratios in the dual noble gases,
this sensor effectively suppresses $B_{\mathrm{gmf}}$ in a common-mode manner while maintaining sensitivity to exotic signals.
In this sensor, the spin precession frequencies of the dual noble gases are measured to extract $\bm{B}_{\mathrm{pseu}}$,
as described by the following equations
\begin{equation}
\begin{aligned}
\label{eq:4}
    \Omega_{1}=&\gamma _{1}\left ( B_{0}+ B_{\mathrm{gmf}}  \right ) -\Omega_{\rm{rot}}+\frac{\mu _{N}  }{\hbar F_{1}  }B_{\mathrm{pseu}} ,\\
    \Omega_{2}=&\gamma _{2}\left ( B_{0}+ B_{\mathrm{gmf}}  \right ) -\Omega_{\rm{rot}}+\frac{\mu _{N} }{\hbar F_{2}  }B_{\mathrm{pseu}} ,\\
\end{aligned}
\end{equation}
where $\gamma _{1}$ ($\gamma _{2}$) and $F_{1}$  ($F_{2}$) represent the gyromagnetic ratios and nuclear spin quantum numbers of the dual noble gas atoms, respectively, and $B_{0}$ is the applied bias field. 
$\Omega_{1}$ and $\Omega_{2}$ are the measured quantities of the spin sensor.
By subtracting these measurements in the form $  \left | \Omega_{1} \right | -\left |R \Omega_{2} \right | $ (where $R=\gamma _{1}/\gamma _{2}$),
we can significantly suppress the interference signal $B_{\mathrm{gmf}}$.
To ensure $B_{\mathrm{pseu}}$ is not diminished, the spin sensor uses atoms with gyromagnetic ratios of opposite signs.
In our prototype, using $^{129}$Xe with $\gamma _{129}=-11.86\,\mathrm{MHz/T}$ and $^{131}$Xe with $\gamma _{131}=3.52\,\mathrm{MHz/T}$ meets this requirement.
Consequently, we can  extract $B_{\mathrm{pseu}}$ as follows
\begin{equation}
 \begin{aligned}
 \label{eq:5}
B_{\mathrm{pseu}}=\frac{  \hbar F_{1} F_{2}\left [ \left ( \left | \Omega_{1} \right | -\left |R \Omega_{2} \right |  \right )-\left ( 1-R \right ) \Omega_{\rm{rot}}
 \right ] 
  }{ \mu _{N}\left ( R F_{1}-F_{2}\right ) }, 
 \end{aligned}  
\end{equation}
where $F_{1}=\frac{1}{2}$ and $F_{2}=\frac{3}{2}$ for $^{129}$Xe and $^{131}$Xe, respectively.
As detailed in the Supplemental Material\,\cite{sup1}, the dual noble-gas sensor can suppress $B_{\mathrm{gmf}}$ by at least 4 orders of magnitude in a common-mode manner.
Thus, the interference noise $B_{\mathrm{gmf}}$ is ultimately reduced to below our search sensitivity.

\begin{figure}[t]
    \centering
    \includegraphics[scale=0.9]{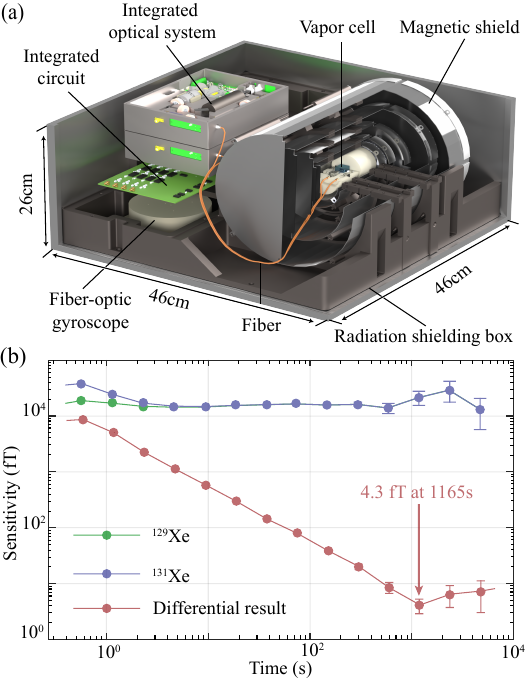}
    \caption{Development of prototype space quantum sensor.
    (a) Design diagram of prototype space quantum sensor.
    The vapor cell containing enriched dual noble gases $^{129}$Xe and $^{131}$Xe, buffer gas N$_{2}$, and a droplet of $^{87}$Rb is placed inside the magnetic shield.   
    The integrated optical system generates both a circularly polarized beam and a linearly polarized beam to pump and probe the spin sensor through fibers.  
    See the Supplemental Material\,\cite{sup1} for more details.
    (b) Allan deviation of the signals for prototype space quantum sensor.
    In the measured signals, the spin precession frequencies of $^{129}$Xe and $^{131}$Xe are read out as $\Omega_{1}$ and $\Omega_{2}$, respectively, while the differential result $  \left | \Omega_{1} \right | -\left |R \Omega_{2} \right | $ is the final out of the sensor. Sensitivity of $\Omega_{1}$ and $\Omega_{2}$ is significantly limited by magnetic noise, whereas this bottleneck can be overcome via dual noble-gas differential.}
    \label{fig:figure 2}
\end{figure}


Figure\,\ref{fig:figure 2}a illustrates the prototype space quantum sensor developed in this work, {which is integrated into a sufficiently small region of $46\times 46\times 26$\,$\mathrm{cm^{3}}$ to meet the payload limitations of the space station\,\cite{spacesd3}}.
Through dual-noble-gas methods, the sensor achieves the best sensitivity of 4.3\,$\mathrm{fT}$ at an integration time of 1165\,s, as shown in Fig.\,\ref{fig:figure 2}b.
The long-term stability and high sensitivity of our spin sensor essentially meet the search requirements of the space scheme for detecting long-period exotic signals.
The sensitivity of the space quantum sensor is determined by various parameters,
including the power and frequency stability of the lasers and the temperature stability of the vapor cell.
Among them, the power stability of the pump laser significantly affects the sensitivity, with a ratio of 19\,$\mathrm{aT}/\mathrm{ppm}$. 
In the experiments, the power stabilization for the integrated optical system in Fig.\ref{fig:figure 2}a achieves 190 ppm at about 1165\,s, as detailed in the Supplemental Material\,\cite{sup1}.
This contributes a noise of 3.7\,$\mathrm{fT}$, which is below the achieved sensitivity of 4.3\,$\mathrm{fT}$.
This highlights the critical role of the laser power stabilization in achieving the current sensitivity and  its potential for further improvements.

The prototype space quantum sensor  could also address other challenges beyond $B_{\mathrm{gmf}}$.
To suppress the noise of $\Omega_{\rm{rot}}$ from space station vibrations, the sensor is deployed with a fiber-optic gyroscope {to measure and subtract the space-station vibration noise $\Omega_{\rm{rot}}$ from Eq.\,\eqref{eq:5}.}
In our test, this gyroscope achieves a sensitivity of $2\times 10^{-6} \,^{\circ} /\mathrm{s}$ within an integration time of about 1165\,s, which could suppress the noise of $\Omega_{\rm{rot}}$ to a negligible value of 0.65\,$\mathrm{fT}$.
Further considering the sensor's disruptions from cosmic radiation, the space quantum sensor is equipped with a 0.5 cm thick aluminum radiation shielding box and implements triple modular redundancy in circuits, ensuring fewer than one disruption per day on average, as detailed in the Supplemental Material\,\cite{sup1}.
This result is acceptable for our searches.
In summary, our experiments demonstrate the practical viability of space-based scheme.

\begin{figure}[t]
    \centering
    \includegraphics[scale=0.88]{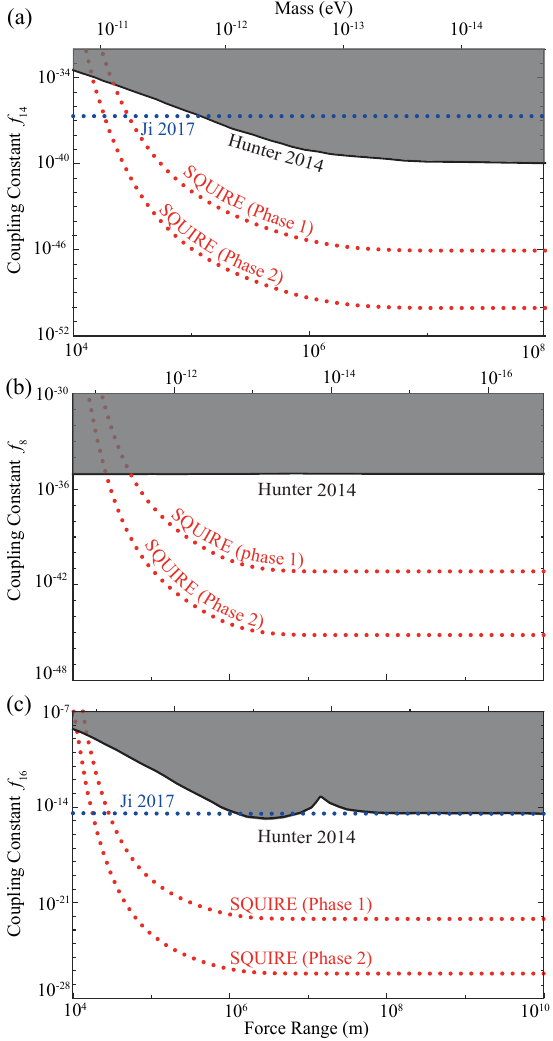}
    \caption{Limits and sensitivity estimations of spin-spin velocity interactions (SSVI).
    For the force range $\lambda>10^4\,$m, the most stringent limits on  SSVI coupling constants are set by the polarized geoelectrons experiment\,\cite{hunter2014using} and proposed by the high-density $\mathrm{SmCo_{5}}$ source experiment\,\cite{ji2017searching1}.
    The constraints and sensitivity estimations are depicted with black solid and blue dotted lines, respectively.
    Our search sensitivity is shown with red dotted lines.
    (a), (b) and (c) show the limits for $V_{14}$, $V_{8}$, and $V_{16}$, respectively.
}
    \label{fig:figure 4}
\end{figure}

\textit{Space-based search sensitivity}.--We estimate the search sensitivity for exotic spin-spin-velocity interactions based on the SQUIRE scheme, dividing it into two phases, both designed to operate under a planned detection duration of 100 days.
Our scheme can improve the search sensitivity for all of the SSVI, including $V_{6+7}$, $V_{8}$, $V_{14}$, $V_{15}$, and $V_{16}$, whose formats provided in the Supplemental Material\,\cite{sup1}.
As shown in Fig.\,\ref{fig:figure 4}, 
our search sensitivity exceeds the sensitivities of both terrestrial experiments\,[\onlinecite{hunter2014using}] and proposals\,[\onlinecite{ji2017searching1}] by more than
7 orders of magnitude for $f_{16}$,
6 orders of magnitude for $f_{14}$, and
6 orders of magnitude for  $f_{8}$ in the force range
$\lambda>10^{6}\,\mathrm{m}$ in phase 1, {which is based on the sensitivity of 4.3\,$\mathrm{fT}$ at an integration time of 1165\,s we achieve.}
Moreover, the search sensitivity for $f_{6+7}$ and $f_{15}$ also exceeds the sensitivities of terrestrial experiments\,[\onlinecite{hunter2014using}] by 4 orders of magnitude in the force range $\lambda>10^{6}\,\mathrm{m}$,
as detailed in the Supplemental Material\,\cite{sup1}.
The six-order-of-magnitude improvement for $f_{14}$ in the space scheme is highly significant and exceptionally challenging to achieve in terrestrial experiments.
For example, to achieve our sensitivity, the proposed high-density $\mathrm{SmCo_{5}}$ source experiment\,\cite{ji2017searching1} (shown as the blue dotted lines in Fig.\,\ref{fig:figure 4}) requires accelerating the velocity of its SmCo5 source to exceed the speed of light,
which is physically impossible.
Furthermore, we also show the phase 2 based on the standard quantum limit of 0.01\,$\mathrm{fT}/\mathrm{Hz}^{1/2}$\,[\onlinecite{kominis2003subfemtotesla}], signifying the ultimate search sensitivity of the scheme.
The search sensitivity is further enhanced by 4 orders of magnitude in phase 2.

In conclusion, we present a space-based scheme to search for spin-spin-velocity interactions mediated by exotic bosons with an unprecedented sensitivity and develop a prototype space quantum sensor capable of overcoming space-environment noise.
Except for the searches of SSVI,
the space quantum sensors can also be extended to search for exotic spin-spin interactions and spin-mass-velocity interactions, including $V_{2}$, $V_{4+5}$ and $V_{12+13}$.
The search sensitivity for those interactions can reach $f_{2}<2\times10^{-51}$, $f_{4+5}<2\times10^{-32}$, and $f_{12+13}<2\times10^{-54}$,
surpassing the existing terrestrial sensitivities by up to 4 orders of magnitude, as detailed in the Supplemental Material\,\cite{sup1}.
Moreover, the space quantum sensor enables highly sensitive searches for axion-halo dark matter surrounding the Earth\,\cite{banerjee2020searching},
greatly outperforming terrestrial experiments.
The coupling between the axion halo and the nuclear spin sensor is proportional to the velocity of the sensor, with $\bm{B}^{\mathrm{DM}}_{\mathrm{pseu}}\propto g_{\phi NN}\bm{v} $,
where $g_{\phi NN}$ is the axion-nucleon spin coupling constant\,\cite{banerjee2020searching}.
The high speed of flying quantum sensors, which exceeds the Earth's rotational linear velocity by an order of magnitude, enables the scheme to achieve a 10-fold improvement in sensitivity for direct dark matter searches compared to terrestrial experiments.

We thank Chengzhi Peng, Lei Zhao, Zhe Cao, Yuanxing Liu, Nan Zhao, Dong Sheng, and Xiangdong Zhang for valuable discussions.
This work was supported by the National Natural Science Foundation of China (Grants Nos.\,T2388102, 92476204, 12274395, 12261160569, and 12404341), the Innovation Program for Quantum Science and Technology (Grant No.\,2021ZD0303205), Youth Innovation Promotion Association (Grant No.\,2023474), the New Cornerstone Science Foundation through the XPLORER PRIZE,
and Frontier Scientific Research Program of Deep Space Exploration Laboratory (Grant No. 2022-QYKYJH-HXYF-013).



\bibliographystyle{naturemag}
\bibliography{mainrefs}

\end{document}